\documentclass[pdflatex,refs]{sn-jnl}

\theoremstyle{thmstyleone}%

\theoremstyle{thmstyletwo}%

\theoremstyle{thmstylethree}%

\usepackage{pifont}
\newcommand{\cmark}{\ding{51}}
\newcommand{\xmark}{\ding{55}}
\usepackage{makecell}
\usepackage{booktabs}
\usepackage{array}
\usepackage{amsmath}
\usepackage{amssymb}
\usepackage{caption}

\usepackage{algorithm}
\usepackage{algpseudocode}

\usepackage{hyperref}

\makeatletter
\let\ps@titlepage\ps@plain
\let\ps@headings\ps@plain
\makeatother
\begin{document}

\title[Article Title]{Graph Symbolic Distillation: Decoding Graph Data into Interpretable Formulas across Materials and Chemistry}

\author[1,2]{\fnm{Yuxuan} \sur{Yao}}  
\author[3]{\fnm{Karsten} \sur{Reuter}} 
\author*[4]{\fnm{Wenbin} \sur{Xu}}\email{wenbin.xu@xmu.edu.cn}

\affil[1]{\orgdiv{Department of Physics and Bavarian Center for Battery Technologies}, \orgname{University of Bayreuth}, \orgaddress{\street{Universitätsstr. 30}, \city{Bayreuth}, \postcode{95447}, \country{Germany}}}

\affil[2]{\orgdiv{Department of Applied Physics and Materials}, \orgname{The Hong Kong Polytechnic University}, \orgaddress{\street{Kowloon}, \city{Hong Kong}, \postcode{999077}, \country{China}}}

\affil[3]{\orgname{Fritz-Haber-Institut der Max–Planck-Gesellschaft}, \orgaddress{\street{Faradayweg 4-6}, \city{Berlin}, \postcode{D-14195}, \country{Germany}}}

\affil*[4]{\orgdiv{College of Energy, State Key Laboratory of Physical Chemistry of Solid Surfaces}, \orgname{Xiamen University}, \orgaddress{\city{Xiamen}, \postcode{361005}, \country{China}}}

\abstract{
Graph data provide a natural representation of the interconnected structures and interactions underlying many chemical and materials systems. 
Graph neural networks (GNNs) can learn complex relationships from such data, yet translating these representations into interpretable descriptors and scientific insights remains challenging.
Here we introduce graph symbolic distillation (GSD), a framework for decoding graph data into interpretable symbolic formulas. 
GSD identifies task-relevant message-passing architectures and distills their learned transformations while preserving graph connectivity and aggregation operations. 
Across three case studies spanning bulk materials, molecules, and surfaces, GSD not only recovers expert-defined descriptors from elementary node and edge features, but also reveals relationships emerging from their organization in complex graphs: extremal local geometric descriptors for band gap variations in $\gamma$-CsPbI$_3$, a local polarity--size relationship underlying hydration free energy, and site-specific electronic--connectivity coupling in multidentate adsorption. 
GSD thus provides a general route from graph data to interpretable descriptors and new, testable scientific insights.
}


\maketitle

\section{Introduction}\label{sec1}

Explainable artificial intelligence (XAI) is emerging as a central paradigm in AI-driven scientific discovery, aiming to move beyond black-box prediction to unveil the governing principles of complex systems.\cite{XAI} 
It seeks to identify physically meaningful descriptors and translate latent space into human-interpretable knowledge.\cite{XAI, gnnsr-cranmer}
Despite this promise, extending XAI to graph-structured data remains particularly challenging. Graphs provide a natural representation for complex systems across chemistry,\cite{gnn-mol} materials science,\cite{CGCNN} biology,\cite{gnn-pharmacy} and social networks,\cite{networks} yet the relationships encoded in graphs are fundamentally different from those in conventional vector-based representations.\cite{battaglia2018} 
In graph-structured data, information is distributed across entities, pairwise interactions, and global variables, with local transformations recursively coupled through permutation-invariant aggregation.\cite{graph-data_2021, graph-data_2022} 
Consequently, graph neural networks (GNNs) and graph kernels encode relationships across nodes, edges, and aggregation pathways, complicating their translation into interpretable rules.

GNNs provide an appropriate inductive bias for graph-structured data by mapping local interactions to system-level properties through message passing (MP).\cite{gnn-1stpaper,gcn} However, the relationships learned through MP remain embedded in a latent numerical space rather than expressed as explicit, explainable scientific rules.
Symbolic regression (SR), in contrast, seeks to represent input--output relationships as explicit mathematical expressions and has emerged as a promising approach to data-driven scientific discovery.\cite{sr-schmidt, sr-makke-review} 
Conventional SR typically uses fixed-length inputs, relying on hand-crafted descriptors or global pooling for graphs, which can obscure topology-dependent information propagation and hierarchical aggregation.

The convergence of graph-based modelling and SR involves both learning from relational data and exploiting structure in the equations themselves.
Hybrid property predictors use SR to refine GNN predictions with selected physical features,\cite{sa-gat-sr} or to enrich descriptor sets and interpret selected latent representations.\cite{mattervial}
These strategies improve prediction and chemical interpretability, although their symbolic components do not reconstruct the full MP computation.
Graph structure also provides a language for equation discovery: GraphED represents candidate constitutive laws as directed graphs, jointly optimizing their algebraic structure and material-specific parameters.\cite{graphed}
For networked dynamical systems, ND$^{2}$ incorporates node--edge transformations and neighborhood aggregation into symbolic search, using neural guidance to recover explicit dynamical laws.\cite{ND2}

A complementary direction extracts symbolic expressions from trained GNN components using sparsity constraints or physical inductive biases.\cite{gnnsr-cranmer,gnnsr-lemos}
Subsequent work made skeleton selection partially data-driven by starting from an overcomplete MP flow and progressively pruning its blocks, layers, connections, and representation dimensions before symbolizing the retained components.\cite{gnnsr-shi} 
These studies demonstrate the potential of extracting explicit symbolic relations from learned GNN components.
For graph-level structure--property prediction, a broader goal is to make the entire predictive process explicit, encompassing both the construction of task-relevant graph descriptors and their mapping to the target property.
The central challenges are therefore to identify a task-appropriate MP flow, and to translate its learned latent representations and nonlinear mappings into physically interpretable relationships.
The latter is particularly difficult in complex scientific systems, where contributions from heterogeneous local environments and coupled interactions can become entangled across latent channels and successive transformations.

Here, we introduce graph symbolic distillation (GSD), which advances graph-based XAI from manually specified or pruning-selected MP skeletons to automated architecture discovery and end-to-end formula extraction. 
GSD combines constrained MP neural architecture search (CMP-NAS), layer-wise symbolic distillation (SD), and hierarchical pruning to obtain analytical models that balance compactness and predictive fidelity. 
These models preserve local interactions and graph-level aggregation, yielding task-adaptive descriptors for physical interpretation and explicit predictors for screening.
Unlike perturbation- or attribution-based explanations that typically focus on individual predictions,\cite{shap,lime} GSD extracts shared symbolic rules from relationships learned across the dataset. 
We demonstrate its generality across bulk, molecular, and surface systems through three case studies: the band gap of $\gamma$-CsPbI$_3$,\cite{metastable} molecular hydration free energies,\cite{freesolv,freesolv-interaction-map} and complex-adsorbate adsorption energies on transition-metal surfaces.\cite{wwl-gpr} 
For $\gamma$-CsPbI$_3$, extremal statistics of local Pb--I bond lengths and out-of-plane orientations describe the strain-dependent band gap. 
For FreeSolv, size-dependent contributions of local polar environments suggest jointly tuning the abundance and distribution of oxygen-containing and hydrogen-bond-donating groups as the carbon framework grows. 
For multidentate adsorption, cross-node electronic couplings, asymmetric coordination effects, and extremal local environments suggest tuning adsorption enthalpy through site-ensemble arrangements.
More broadly, GSD provides a general route for AI for science to move beyond prediction towards automated scientific abstraction.

\section{Results}\label{sec2}

\subsection{Overview of Graph Symbolic Distillation (GSD)}
\label{subsec2-1}

\begin{figure*}[htp]
\centering
\includegraphics[width = \textwidth]{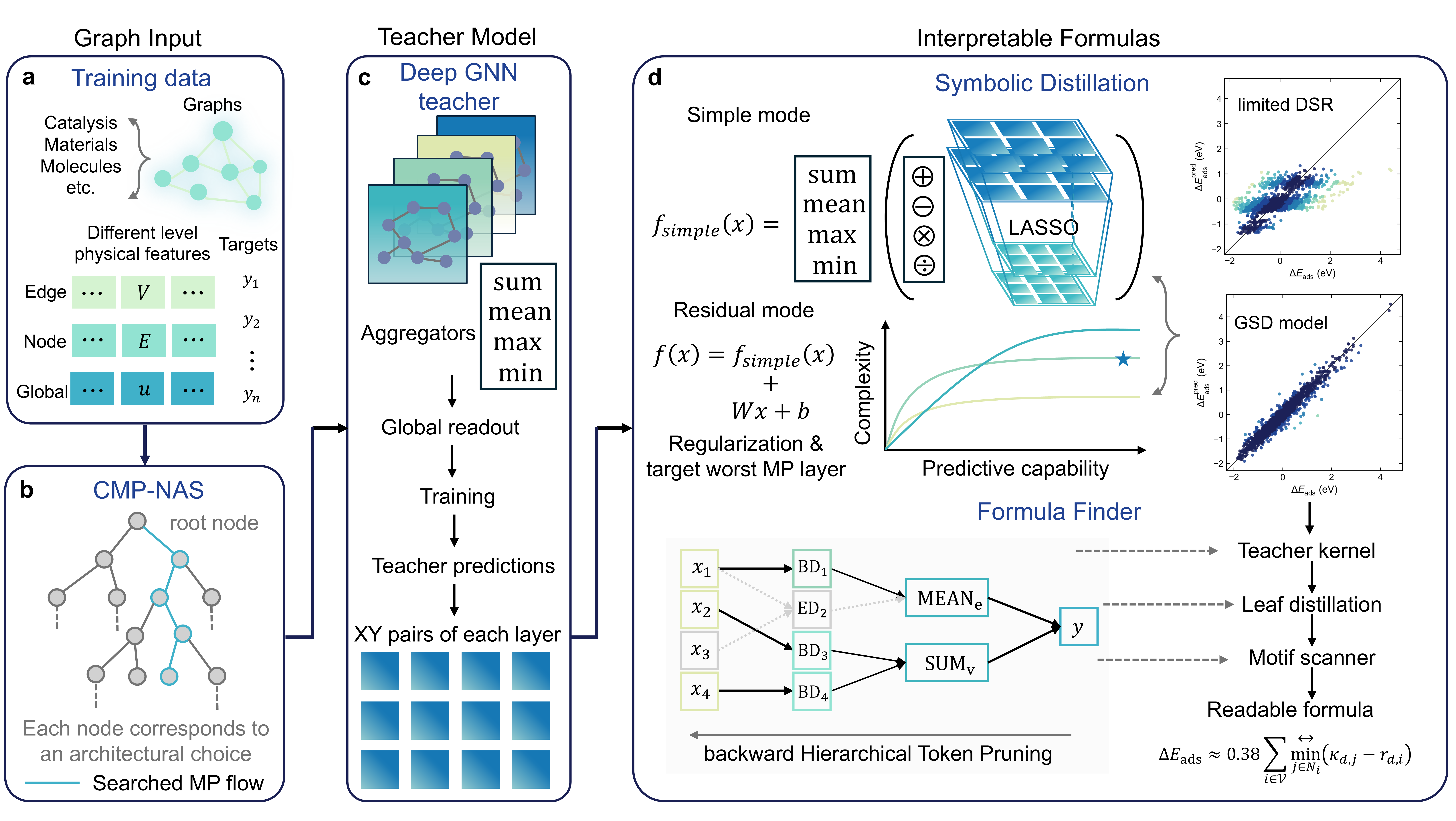}
\captionsetup{font=footnotesize,skip=2pt}
\caption{\label{fig:scheme} Schematic overview of GSD. 
(a) Graph-structured data are first used to search for a task-specific MP flow through (b) CMP-NAS (See Section S3 of the SI and Methods for further details). 
(c) The selected flow is instantiated as a neural teacher, from which layer-wise input--output pairs are extracted for symbolic fittings.
(d) SD then replaces the learned neural transformations with symbolic expressions using either the simple or residual mode, followed by formula finder workflow, integrating hierarchical symbolic-token pruning with coefficient refinement, to obtain the physical formula (See Section S5 of the SI and Methods for further details).
}
\end{figure*}

A predictive GNN can be conceptually decomposed into two core components: graph representation learning, which constructs representations through hierarchical message passing; and property mapping, which maps these learned representations to target properties.
These components give rise to two central challenges in interpreting graph-based models. First, capturing task-relevant relationships from graph-structured data requires an appropriate MP architecture, whose design is inherently task-dependent. 
Second, the resulting predictions rely on latent representations and nonlinear transformations, making it difficult to translate the learned graph representations into explicit, human-interpretable formulas.
GSD addresses the architecture-design challenge through CMP-NAS, which automatically discovers how information is transformed, propagated, and aggregated for a given task. 
For the interpretability challenge in property mapping, SD and a formula finder workflow translate the learned transformations into compact formulas that link input features to predictions.
These two objectives are realized through the four-stage framework shown in Fig.~\ref{fig:scheme}: MP-flow search, teacher-model training, SD, and formula construction.

First, as shown in Figs.~\ref{fig:scheme}a,b, CMP-NAS uses constrained Monte Carlo tree search (MCTS) to select an MP flow, specifying its depth, edge and node updates, aggregation operators, and global readout. This computational structure remains fixed throughout the subsequent stages.
Second, the selected architecture is trained as a neural teacher to predict the target property. Input--output pairs extracted from its intermediate update functions then provide supervision for SD (See Fig.~\ref{fig:scheme}c).
Subsequently, SD, schemed in Fig.~\ref{fig:scheme}d, uses SR to approximate each scalar output channel of the teacher’s update functions with a sparse symbolic expression while preserving the MP flow and its aggregation operators. The distilled expressions are either purely symbolic or supplemented by linear residual corrections, after which all coefficients are jointly optimized against the target property. 
Note that the component-wise fitting of intermediate mappings distinguishes SD from direct symbolic regression (DSR), which maps fixed-length descriptors directly to the target property. All symbolic regressions in this work are performed using TorchSISSO,\cite{torchsisso} which implements the sure independence screening and sparsifying operator (SISSO),\cite{sisso} a compressed-sensing-based method widely used for descriptor discovery and interpretable modeling in chemistry and materials science.\cite{sisso-catalysis,sisso-luca,sisso-Octahedralmotif}
Finally, the formula finder applies backward hierarchical token pruning to remove terms with negligible predictive contributions, thereby enhancing interpretability. It then composes the retained expressions along the MP flow and expands them in terms of the original input features, yielding a compact formula with graph aggregations (See Fig.~\ref{fig:scheme}d).
Owing to this end-to-end framework, GSD thereby transforms the black-box graph-to-property prediction into an interpretable graph-to-descriptor-to-law framework.

In the following sections, we describe two key aspects of GSD's design, namely CMP-NAS and SD, and evaluate the GSD framework against DSR on three graph-regression tasks (Figs.~\ref{fig:case1}--\ref{fig:case3}): the band gap $E_{\mathrm{g}}$ of distorted $\gamma$-CsPbI$_3$ crystals (Fig.~\ref{fig:case1}), the molecular hydration free energy $\Delta G_{\mathrm{hyd}}$ in FreeSolv (Fig.~\ref{fig:case2}), and the adsorption enthalpy $\Delta E_{\mathrm{ads}}$ of complex adsorbates on catalytic surfaces (Fig.~\ref{fig:case3}). 
Their complete feature definitions and graph-construction procedures are provided in Section S1 and Tables S1–S3 of the Supporting Information (SI). 
Further details on the notation used in the GSD framework, the CMP-NAS procedure, the symbolic regression setup, and the formula-finder workflow are provided in Methods.


\subsection{Constrained MP Neural Architecture Search (CMP-NAS) Strategy}
\label{subsec2-2}

\begin{figure*}[!htp]
\centering
\includegraphics[width = 1 \textwidth]{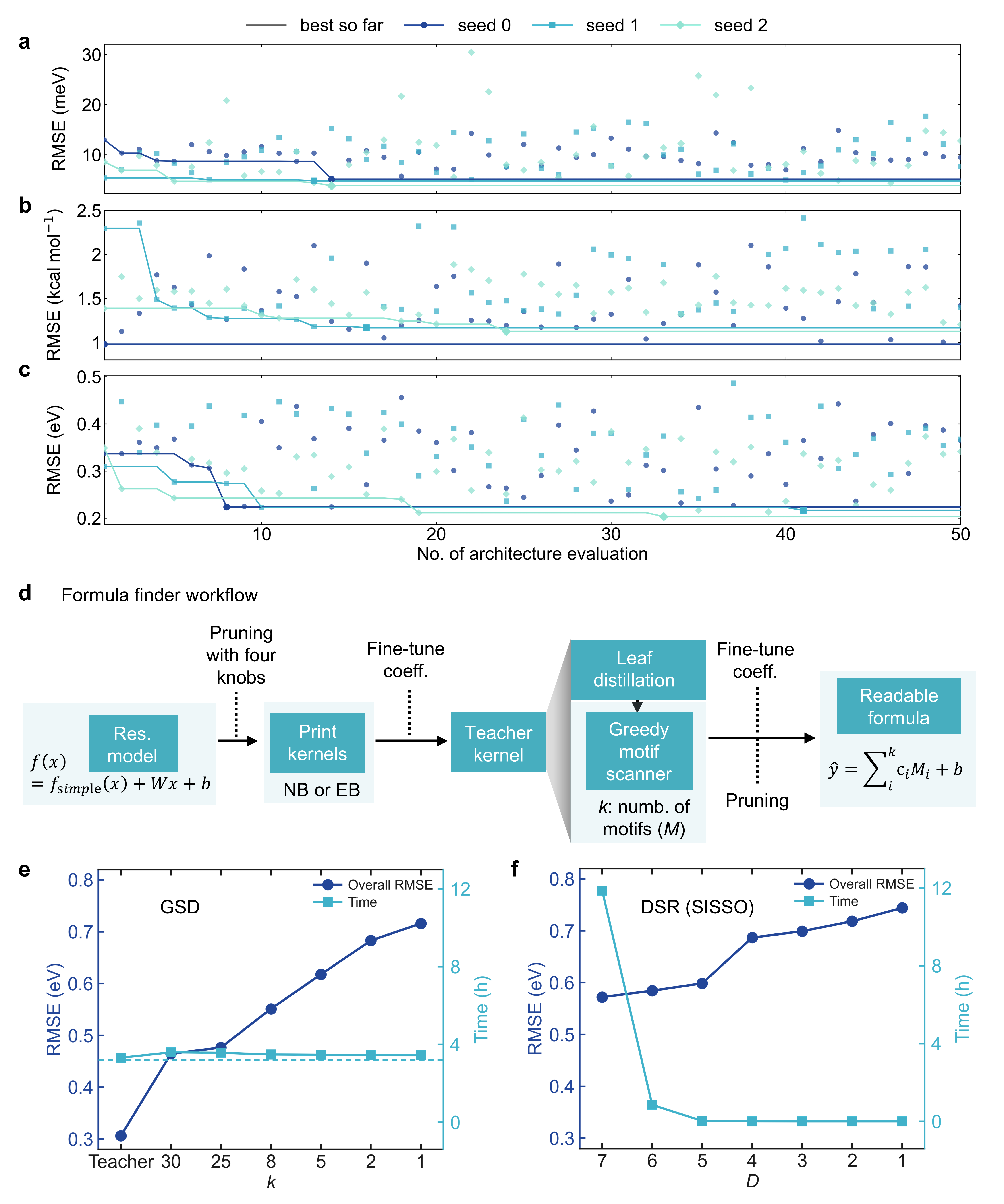}
\captionsetup{font=footnotesize,skip=2pt}
\caption{\label{fig:mcts} 
Architecture search and formula distillation in GSD. 
Convergence of the MCTS-based CMP-NAS for (a) $\gamma$-CsPbI$_3$, (b) FreeSolv and (c) ComplexAdsorbate tasks. Each point represents the validation RMSE obtained from one architecture evaluation, in which a completed MP flow is instantiated, trained, and evaluated. Solid lines show the lowest validation RMSE found up to the corresponding evaluation. Results from three independent random seeds are shown.
(d) Formula-finder workflow, showing pruning, coefficient refinement, and motif extraction from the distilled model.
(e,f) Predictive accuracy and computational cost as a function of formula complexity for the ComplexAdsorbate task: GSD with varying numbers of retained graph motifs \(k\), and direct symbolic regression using SISSO with varying descriptor dimensions \(D\).
Teacher denotes the teacher kernel in (d), and the cyan dashed line marks its time at 3.31 hours, which is measured from the end of Res. model to the start of the greedy search, comprising the print kernels step (1.18 h) and the teacher kernel finetune step (2.13 h).
}
\end{figure*}

In GSD, an explicit MP flow organizes intermediate transformations into input--output pairs for layer-by-layer distillation.
Its aggregation operators are also structurally informative: sum pooling preserves extensive contributions, mean pooling describes size-normalized environments, and maximum and minimum pooling retain extremal local responses.
Identifying the MP flow and its aggregation operators is therefore a prerequisite for SD.

The search space of MP flow is derived from the standard graph-network formalism of Eqs.~\ref{eq:edge-update}--\ref{eq:global-update}.\cite{battaglia2018}
Combinations of computational components, learnable transformations $\phi$, and aggregation operators $\rho$ yields a vast architecture space, making the identification of a task-appropriate MP flow computationally challenging.
To make this search tractable while preserving the typed structure of graph representations, we developed CMP-NAS, which restricts candidates to the bottom-up slot template in Eq.~\ref{eq:mpflow_template}, and selects only operations compatible with the edge-, node-, or graph-level representation available at each slot. 
We then employ the MCTS-based method to explore this constrained space by progressively constructing MP flows. Each partial flow is completed with a uniformly sampled legal suffix, and the terminal architecture is evaluated by its root mean square error (RMSE) under coarse single-split validation.\cite{uniform}
Full details of the MCTS-based algorithm are provided in Methods.\cite{uct,mcts}

This uniform-completion strategy used in CMP-NAS simplifies the search while introducing additional variance into the architecture rewards. Thus, CMP-NAS exhibits task-dependent convergence behavior, as illustrated by searches conducted with three random seeds (see Figs.~\ref{fig:mcts}a--c).
For $\gamma$-CsPbI$_3$ and FreeSolv tasks (Figs.~\ref{fig:mcts}a,b), multiple seeds identify the lowest-error architecture within the first few evaluations, after which the best-so-far curve keeps nearly flat.
These early plateaus are consistent with favorable terminal completions.
By contrast, the harder complex adsorbates task (Fig.~\ref{fig:mcts}c) shows more frequent stepwise improvements, with better architectures appearing later despite substantial reward dispersion.

We next evaluated the predictive performance of the MP architectures identified by CMP-NAS across the three graph-learning tasks using five-fold cross-validation (see Table ~\ref{tab:literature_comparison}).
Compared with reported graph-based baselines, the identified architectures achieve a 92\% higher MAD/MAE ratio for $\gamma$-CsPbI$_3$, an 11\% lower MAE for FreeSolv, and a 41\% lower RMSE for ComplexAdsorbate.
Despite data splits and training protocols may differ, these comparisons provide evidence of the strong predictive performance of the identified architectures. More importantly, the identified architectures support SD into explicit formulas detailed in Section 2.4. For each task, we select the best-performing MP flow across the three search runs and use the same random seed for subsequent stages.

\begin{table}[t]
\centering
\renewcommand{\arraystretch}{1.2}
\captionsetup{
  justification=justified,
  singlelinecheck=false
}
\caption{Predictive performance of CMP-NAS-searched MP models compared with
previously reported graph-based models. All comparison models were evaluated
using the same task-specific input feature sets. The only modification was made
for the ComplexAdsorbate task, where the SOAP descriptors used in the original 
implementation were excluded to retain a physically interpretable input feature space.}
\label{tab:literature_comparison}

\resizebox{\linewidth}{!}{%
\begin{tabular}{c c c l l c}
\toprule
Dataset
& Size
& Target
& Model
& Performance
& \makecell[c]{Explicit formula} \\
\midrule

\multirow{2}{*}{$\gamma$-CsPbI$_3$}
& \multirow{2}{*}{505}
& \multirow{2}{*}{\makecell[c]{$E_{\mathrm g}$\\(eV)}}
& CGCNN~\cite{metastable}
& MAD/MAE = 9.01
& \xmark \\

& & &
This work
& \textbf{MAD/MAE = 17.27}
& \cmark \\

\midrule

\multirow{2}{*}{FreeSolv}
& \multirow{2}{*}{642}
& \multirow{2}{*}{\makecell[c]{
    $\Delta G_{\mathrm{hyd}}$\\
    (kcal mol$^{-1}$)
}}
& GNN~\cite{freesolv-interaction-map}
& MAE = $0.70 \pm 0.10$
& \xmark \\

& & &
This work
& \textbf{MAE = 0.62}
& \cmark \\

\midrule

\multirow{2}{*}{\makecell[c]{Complex\\Adsorbate}}
& \multirow{2}{*}{1,679}
& \multirow{2}{*}{\makecell[c]{
    $\Delta E_{\mathrm{ads}}$\\
    (eV)
}}
& WWL--GPR(no SOAP)~\cite{wwl-gpr}
& RMSE = 0.32
& \xmark \\

& & &
This work
& \textbf{RMSE = 0.19}
& \cmark \\

\bottomrule
\end{tabular}%
}
\end{table}


\subsection{Symbolic Distillations (SD) and its Fidelity--Complexity Trade-offs}\label{subsec2-3}

Once the identified MP flow has been trained as a deep teacher, GSD approximates its learned transformations through layer-wise SD. 
GSD has two primary modes: GSD–Simple (termed Simp.), which replaces the learned transformations with purely symbolic expressions, and GSD–Residual (termed Res.), which augments these expressions with linear corrections to enhance expressive capacity (Eq.~\ref{eq:sr-layer-wise} and Eq.~\ref{eq:sr-residual}). 
In order to control model complexity, we further consider two residual variants. 
GSD--$\ell_1$ Residual (termed $\ell_1$ Res.) applies an $\ell_1$ penalty of $\lambda=0.001$ to all residual coefficients to promote sparsity.\cite{lasso}
GSD--Targeted Residual (termed Targ.) restricts the correction to the MP component with the largest symbolic-fitting error. 
Each of the four variants yields a distilled model that preserves the selected MP topology and aggregation operators.

Expressions with fewer terms and dependencies are generally easier to inspect and relate to physical mechanisms. 
Although additional model capacity can improve the approximation of the teacher’s transformations, it does not necessarily improve predictive performance on unseen data. 
For each task, Table S4 compares the four GSD variants with DSR and the deep teacher GNN. To characterize model complexity, trainable parameters describe the student before pruning, whereas formula complexity counts the symbolic coefficients and residual linear terms retained afterward. 
This table reveals a task-dependent trade-off between predictive fidelity and model complexity. For $\gamma$-CsPbI$_3$ and ComplexAdsorbate, Res. achieve the lowest test RMSE among the distilled models, but retain the most complex formulas. 
Specifically, for $\gamma$-CsPbI$_3$, Targ. reduces formula complexity from 397 to 304, yet with test RMSE slightly increasing from 0.00269 to 0.00490~eV.
Similarly, for ComplexAdsorbate, $\ell_1$ Res. reduces complexity from 1,912 to 801, with a smaller relative increase in test RMSE from 0.23 to 0.27~eV. 
For FreeSolv, however, Targ. achieves achieves a lower test RMSE than Res. (1.16 vs. 1.40~kcal~mol$^{-1}$) with substantially fewer trainable parameters (389 versus 3,159), while maintaining comparable formula complexity (220 versus 219). 

Figs.~\ref{fig:case1}--\ref{fig:case3}f show the trade-off between predictive error and model complexity across variants.
The deep teacher GNNs generally achieve the lowest prediction errors but require substantially greater model complexity, whereas the distilled models offer more compact symbolic representations with varying levels of predictive fidelity. 
Notably, Targ. for $\gamma$-CsPbI$_3$ and FreeSolv, and $\ell_1$ Res. for ComplexAdsorbate, provide favorable balances between predictive accuracy and symbolic complexity. These variants are therefore selected for subsequent formula construction  (see Figs.~\ref{fig:case1}d, \ref{fig:case2}d, and \ref{fig:case3}d).

The GSD models, particularly the Simp. and Targ. variants, already achieve substantial reductions in formula complexity, providing a basis for the direct interpretation or further exploration assisted by model analysis tools, ranging from conventional sensitivity analysis\cite{sensitivity} to more interactive LLM-based assistants.\cite{talktomodel} 
To obtain shorter expressions whose physical meaning is readily apparent to the reader, the formula finder (Fig.~4d) further compresses the selected GSD models into compact, task-specific formulas through hierarchical token pruning and coefficient refinement.
This procedure removes weakly weighted terms and unused dependencies while retaining explicit feature combinations and graph aggregation operations. The resulting predictor is expressed as an additive combination of $k$ graph motifs constructed from the original input features (Eq.~\ref{eq:formula_finder_outer}), making the learned structure--property relationships directly traceable. 
The retained motif count $k$ controls formula complexity, analogous to the descriptor dimension $D$ in SISSO: both specify the number of additive components, although these components differ in construction. 
Further details of the formula finder workflow are provided in Methods and Section~S5 of the SI, with a detailed comparison of $k$ and $D$ in Methods.

\subsection{Applications and New Scientific Insights Uncovered by GSD}
\label{subsec2-4}

Having established the GSD framework and evaluated its predictive fidelity and symbolic complexity, we next examine how the resulting formulas expose interpretable relationships in graph-structured data and suggest testable scientific hypotheses.
Through three case studies spanning materials, molecules, and surfaces, we investigate how GSD uncovers extremal local geometries governing band gap variations, derives molecular design hypotheses from local polarity--size relationships, and reveals site-specific electronic--connectivity coupling in multidentate adsorption.

\begin{figure*}[htp]
\centering
\includegraphics[width = 1 \textwidth]{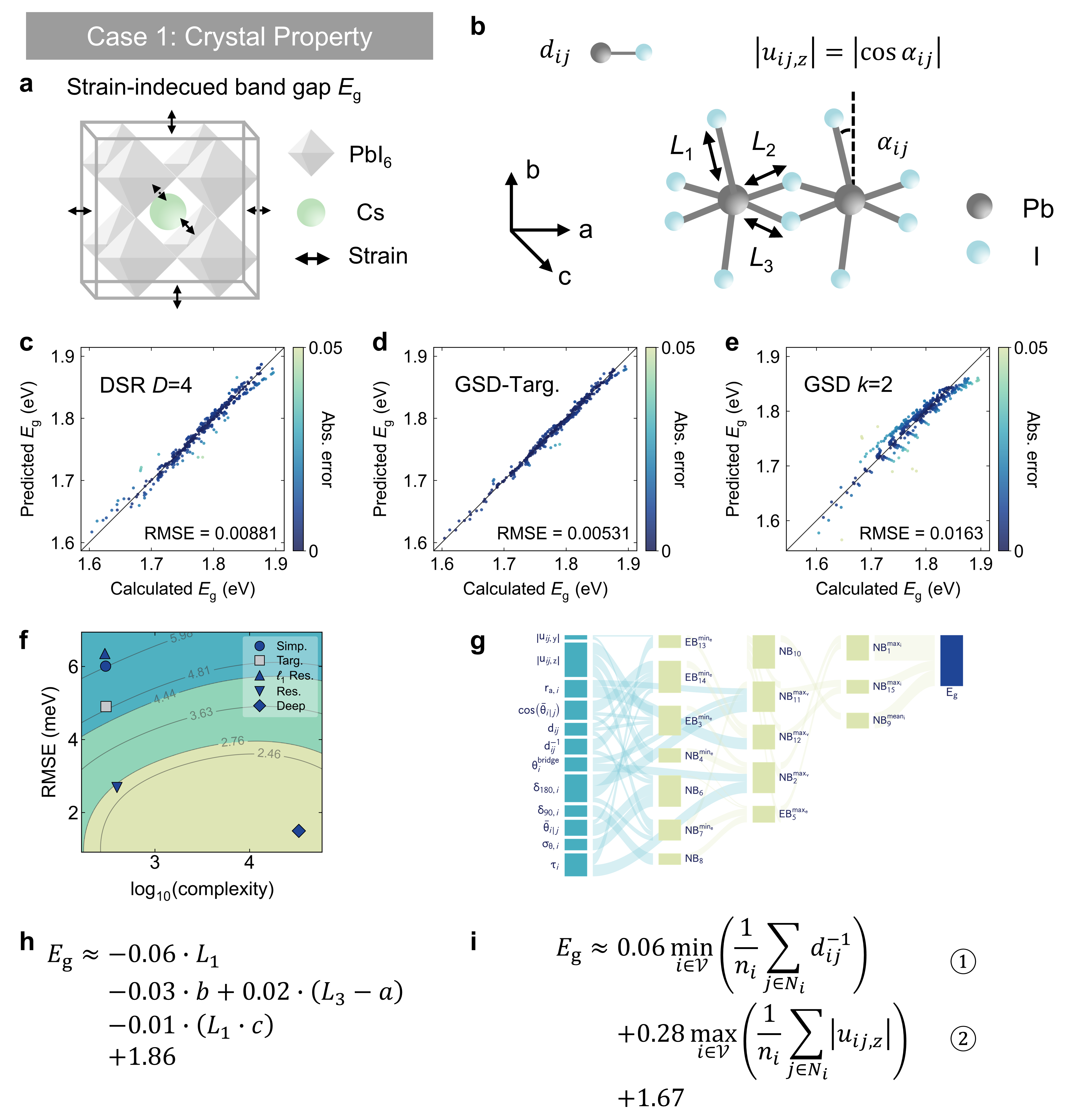}
\captionsetup{font=footnotesize,skip=2pt}
\caption{
Graph-based symbolic modeling of the strain-induced band gap $E_{\mathrm{g}}$ in $\gamma$-CsPbI$_3$.
(a) Schematic of the prediction task and (b) selected structural features.
(c,d,e) Parity plots for DSR formula in (h), GSD--Targ. model, and the compact GSD formula in (i), respectively.
In (i) and also throughout all three cases, $\mathcal{V}$ denotes the node set, $N_i$ the neighbor set of node $i$, and $n=\lvert \mathcal{V} \rvert$ and $n_i=\lvert N_i \rvert$ the corresponding numbers of nodes and neighbors, respectively. 
(f) Performance--complexity landscape. Markers denote Simple (circle), Targeted Residual (square), $\ell_1$ Residual (upward triangle), Residual (downward triangle), and the deep teacher (diamond); contours indicate interpolated iso-RMSE levels.
(g) Sankey diagram of the selected Targ. model, linking input features to $E_{\mathrm{g}}$ through intermediate symbolic expressions NB and EB. Ribbon widths represent relative coefficient-weighted contributions within each stage; superscripts denote aggregation operators.
The landscape reports test-set RMSE, whereas the parity plots report overall RMSE evaluated over the combined training, validation, and test sets. 
This convention applies throughout the work unless otherwise specified.
}
\label{fig:case1}
\end{figure*}

\textbf{$\gamma$-CsPbI$_3$ case: Uncovering Extremal Local Geometry in Band Gap Variations}

The band gap ($E_\mathrm{g}$) of metal halide perovskites determines their optical absorption edge, and is important for single-junction or tandem solar cells.\cite{Perovskite2015, perovksite2016} 
Understanding its dependence on structural distortions is therefore essential for tuning their optoelectronic properties.\cite{bianMLP2026} 
Existing studies have widely used hand-crafted structural descriptors to characterize structure–property relationships in halide perovskites.\cite{pbipb-2014,pbipb2025} 
Here, we use a dataset of strained $\gamma$-CsPbI$_3$ structures to investigate whether GSD can identify relevant structural descriptors directly from graphs and quantify their contributions to $E_\mathrm{g}$. \cite{metastable}
Conventional DSR relies on expert-defined tabular features, with its resulting formula, shown in Fig.~\ref{fig:case1}h, combining lattice parameters (i.e. $a$ and $c$) and Pb--I bond lengths (i.e. $L_1$ and $L_3$). 
In contrast, GSD takes elementary node and edge features as input, without using these tabular descriptors, and automatically constructs interpretable structural descriptors through MP and aggregation. 
For comparison, we also apply GSD using only the same tabular features used by DSR, treating them as global features without any node or edge inputs. The distilled formula selects $L_1$, $L_2$, and $c$, sharing $L_1$ and $c$ with the DSR expression (SI Section~S4). Note that this comparison is separate from the GSD model presented in Fig.~\ref{fig:case1}.

The graph-derived GSD formula reveals complementary bond-length and orientational contributions to $E_\mathrm{g}$ (see Figs.~\ref{fig:case1}b,i). 
Term (1) recovers a bond-length dependence consistent with the negative $L_1$ terms in DSR (Fig.~\ref{fig:case1}h): increasing Pb--I distances lowers the inverse-distance ($d_{ij}^{-1}$) contribution. This descriptor is constructed directly from the atomic graph by averaging inverse Pb--I bond lengths over the neighbors of each node and taking the minimum across all nodes. 
Term~(2) captures out-of-plane framework distortion through the bond-orientation feature $\lvert u_{ij,z} \rvert=\lvert \cos\alpha_{ij} \rvert$, where $\alpha_{ij}$ denotes the angle between the Pb--I bond and the fixed $z$ axis. Its neighborhood average characterizes local out-of-plane reorientation, while the outer maximum selects the environment with the strongest mean out-of-plane alignment.\cite{pbipb2025} 
The positive coefficient associates a larger value of this descriptor with a wider predicted band gap. 

The recovered orientational dependence is consistent with the original study, which found that incorporating local angular information substantially improves predictions for strained $\gamma$-CsPbI$_3$.\cite{metastable}
The use of extremal geometry features also echoes previous findings that the largest metal--halide--metal bond angle serves as an effective band gap descriptor.\cite{pbipb-2014} 
GSD identifies two complementary extremal neighborhood descriptors that capture local bond-length scales and directional alignment. This suggests that selected local geometric environments, rather than global averages alone, may provide informative descriptors of band gap variations. 
Their linear combination with an intercept predicts $E_\mathrm{g}$ across the sampled volume-preserving strains with an RMSE of 0.0163~eV (Fig.~\ref{fig:case1}e). 
This complementarity arises from the formula finder's leaf distillation and motif scanner procedures, which retain an additional leaf only if it improves predictive performance beyond the existing expression (see Methods).

\begin{figure*}[!htbp]
\centering
\includegraphics[width = 1 \textwidth]{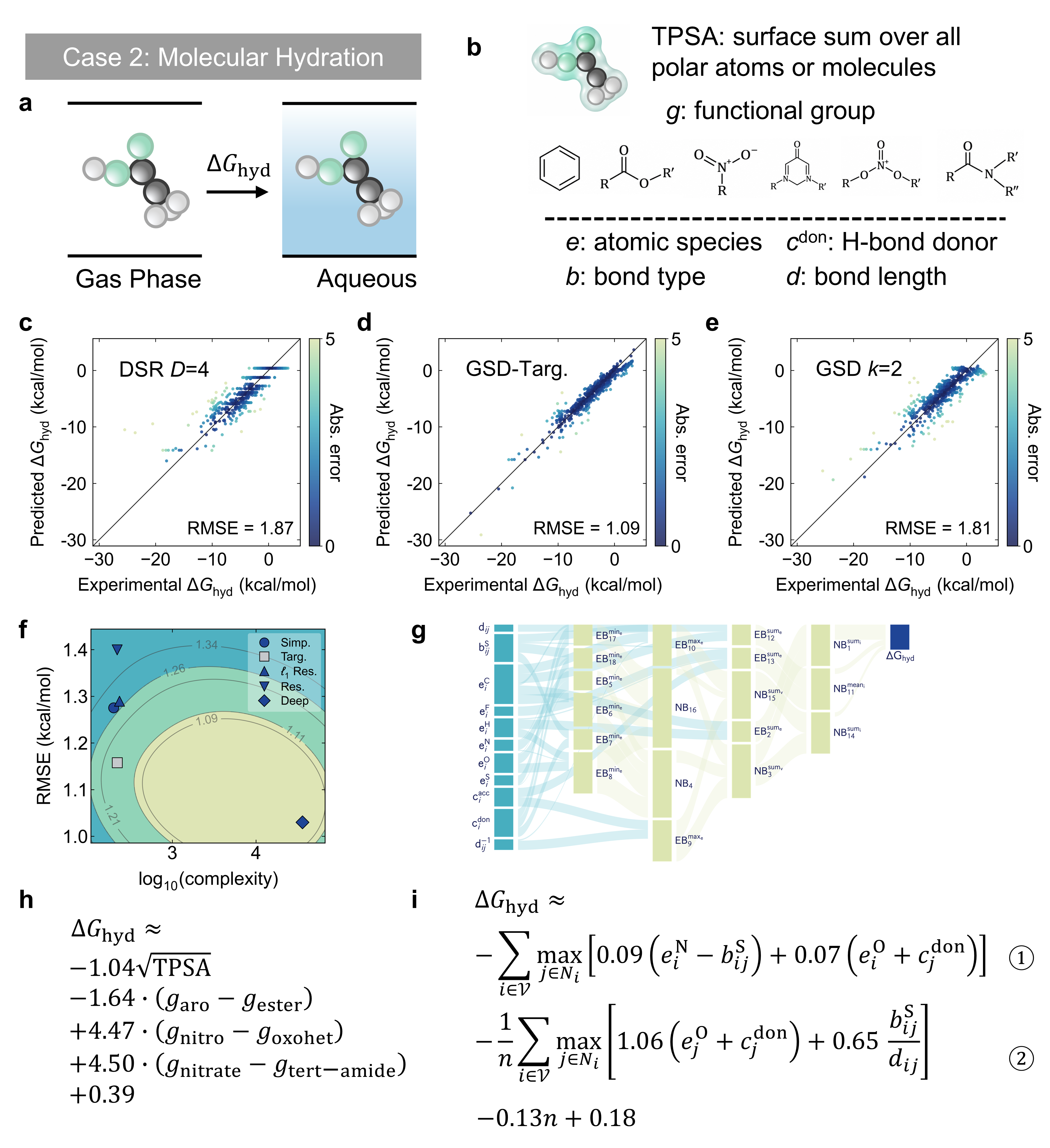}
\captionsetup{font=footnotesize,skip=2pt}
\caption{
Graph-based symbolic modeling of the experimental hydration Gibbs free energy $\Delta G_{\mathrm{hyd}}$ in FreeSolv.
(a) Schematic of molecular hydration and (b) representative global, node, and edge features.
(c,d,e) Parity plots for DSR formula in (h), GSD Targ. model, and the GSD formula in (i), respectively.
(f) Performance--complexity landscape, with model symbols defined as in Fig.~\ref{fig:case1}f and contours indicating interpolated iso-RMSE levels.
(g) Sankey diagram of the selected Targ. model. Ribbon widths represent relative coefficient-weighted contributions through intermediate expressions NB and EB; superscripts denote aggregation operators.
}
\label{fig:case2}
\end{figure*}

\textbf{FreeSolv: Revealing Local Polarity--Size Relationships to Guide Molecular Design}

The hydration free energy ($\Delta G_{\mathrm{hyd}}$) quantifies the free-energy change associated with transferring a molecule from the gas phase into water and is a fundamental thermodynamic property relevant to solvation and molecular design, such as drug discovery.\cite{freesolv,solv-energy} 
Understanding how molecular composition and local bonding environments influence ($\Delta G_{\mathrm{hyd}}$) is essential for identifying structural modifications that favor hydration.\cite{localbonding} 
However, conventional molecular descriptors often summarize chemical features at the molecular level, making it difficult to distinguish the roles of individual atomic environments in the overall structure--property relationship.\cite{local1,local2} 
To examine whether GSD can uncover these relationships directly from molecular graphs, we consider the FreeSolv dataset of small neutral molecules.\cite{freesolv} 
Using predefined global molecular descriptors,\cite{freesolv-interaction-map} conventional DSR yields a formula combining topological polar surface area (TPSA) and functional-group indicators (Figs.~\ref{fig:case2}b,h). 
In contrast, graph-based GSD automatically constructs two chemically interpretable descriptors from elementary atomic and bond features, together with an explicit molecular-size correction (Figs.~\ref{fig:case2}b,i).

The graph-derived GSD formula reveals complementary contributions from local chemical environments and molecular size. 
Term (1) sums local maximum scores involving nitrogen and oxygen identity ($e^{N}$ and $e^{O}$) and hydrogen-bond-donor character ($c^\mathrm{don}$), modulated by the single-bond indicator ($b^\mathrm{s}$).  
Term (2) averages local maximum scores combining oxygen and donor indicators ($e^{O}$ and $c^\mathrm{don}$) with single-bond-weighted inverse bond lengths ($b^{\mathrm S}/d$).
These distinct graph aggregation operations capture the molecular accumulation of chemically weighted local environments in Term (1) and the average strength of local chemical features per atom in Term (2). 
The remaining $-0.13n$ term introduces an explicit dependence on the number of atoms.

The recovered dependence of hydration free energy ($\Delta G_{\mathrm{hyd}}$) on heteroatom identity and hydrogen-bond-donor character is consistent with the favorable hydration of polar molecular environments, with larger corresponding scores associated with more negative $\Delta G_{\mathrm{hyd}}$.\cite{polar,freesolv-interaction-map} 
The single-bond indicator and inverse bond-length features further distinguish local bonding environments, reflecting how heteroatoms are connected within the molecular graph. This distinction is chemically relevant because nitrogen hydrogen-bond acceptor strength depends on hybridization and electronic environment\cite{nitrogen-hbond-2018}, while donor and acceptor strengths vary across functional groups in organic molecule–water complexes.\cite{water-hbond-1998} 

Beyond recovering these individual chemical relationships, GSD reveals how they combine through distinct graph aggregation operations, suggesting a more specific molecular design rule: as the carbon framework grows, introducing oxygen-containing or hydrogen-bond-donating groups across the molecular scaffold may favor hydration by increasing the cumulative polar contributions while maintaining their average strength per atom. 
Local maximum aggregation further suggests that adding a polar group to a weakly polar neighborhood may be more effective than adding one that leaves an existing neighborhood maximum unchanged. Together, these relationships suggest that favorable hydration may depend not only on the number of polar functional groups but also on their local distribution and balance with molecular size, providing a testable hypothesis for molecular design.

The resulting compact formula achieves an overall RMSE of 1.81~kcal~mol$^{-1}$, comparable to 1.87~kcal~mol$^{-1}$ for DSR (Fig.~\ref{fig:case2}c,e). 
By revealing how local polar environments and molecular size are jointly associated with hydration free energy, GSD provides an interpretable structure--property relationship and generates hypotheses for molecular design based on functional-group identity, placement, and scaffold size.

\begin{figure*}[!htbp]
\centering
\includegraphics[width = 1 \textwidth]{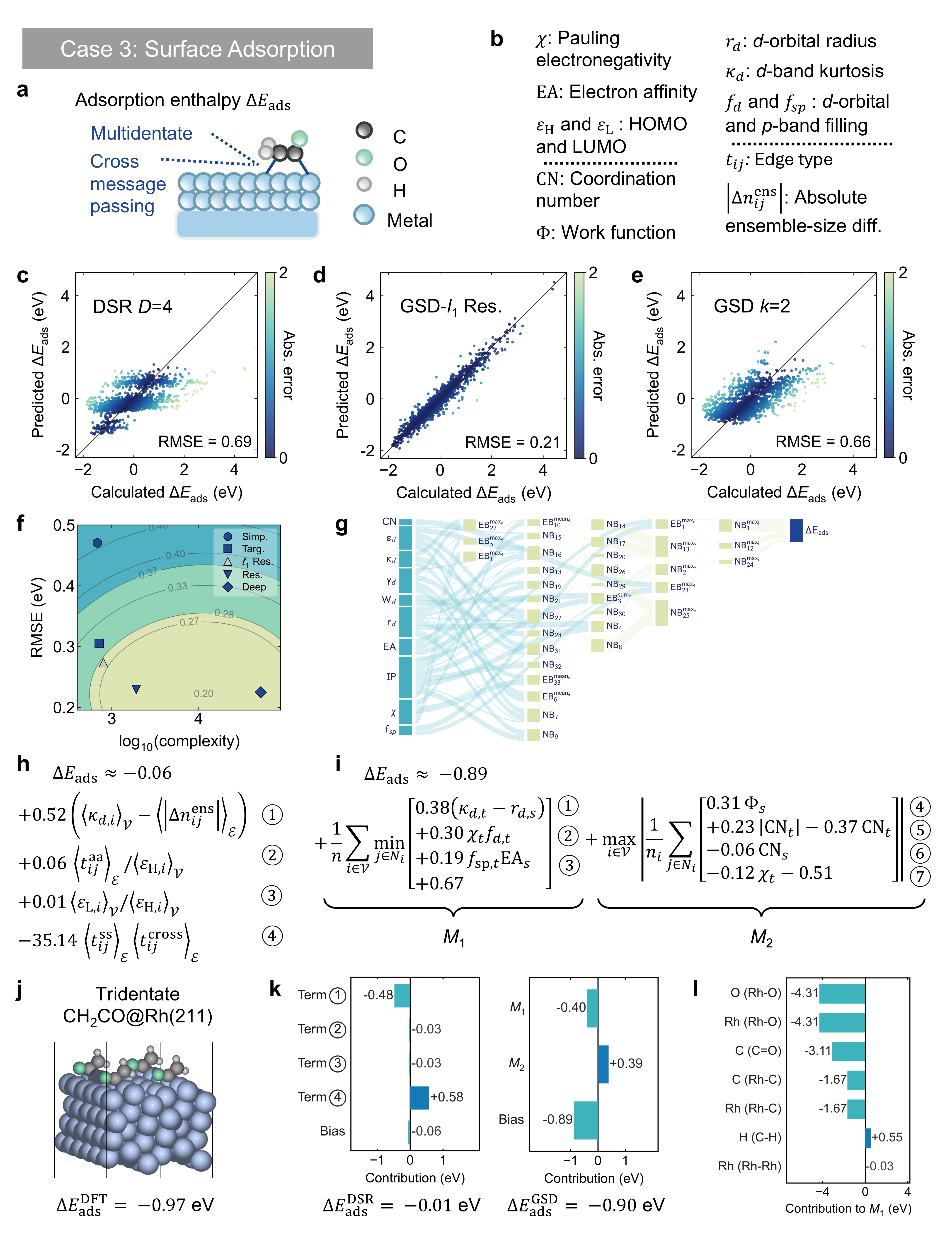}
\captionsetup{font=footnotesize,skip=2pt}
\caption{
Graph-based symbolic modeling of the adsorption enthalpy $\Delta E_{\mathrm{ads}}$ in complex adsorbates.
(a) Schematic of  multidentate adsorption across the adsorbate--surface interface and (b) representative node, and edge features.
(c,d,e) Parity plots for DSR formula in (h), GSD $\ell_1$ Res. model, and the GSD formula in (i), respectively. 
In (h), mean-pooling node and edge attributes is denoted by $\langle\cdot\rangle_{\mathcal{V}}$ and $\langle\cdot\rangle_{\mathcal{E}}$, respectively. 
In (i), $s=\min(i,j)$ and $t=\max(i,j)$ by atom index. Metal atoms are indexed before adsorbate atoms in each graph. 
(f) Performance--complexity landscape, with model symbols defined as in Fig.~\ref{fig:case1}f and contours indicating interpolated iso-RMSE levels.
(g) Sankey diagram of the selected Targ. model. Ribbon widths represent relative coefficient-weighted contributions through intermediate expressions NB and EB; superscripts denote aggregation operators.
(j) Tridentate CH$_2$CO adsorption on Rh(211) taken as an example on which the GSD formula succeeds, with a DFT adsorption enthalpy of $-0.97$~eV.
(k) Contributions of individual components and the intercept to the DSR (left) and GSD (right) predictions for (j); the final bar gives their sum.
(l) Atom-resolved neighborhood minimum scores entering $M_1$ in (i). Their arithmetic mean gives $M_1$.
}
\label{fig:case3}
\end{figure*}

\textbf{ComplexAdsorbate: Decoding Site-Specific Electronic--Connectivity Coupling in Multidentate Adsorption}

The adsorption enthalpies ($\Delta E_{\mathrm{ads}}$) of complex adsorbates, such as multicarbon intermediates, carboxylates, and other larger molecular fragments, are key thermodynamic quantities in heterogeneous catalysis, relevant to industrially important processes such as Fischer–Tropsch synthesis,\cite{ftsynthesis} CO$_2$ reduction,\cite{ccco2} and methanol\cite{methanol1, methanol2} and higher oxygenate synthesis.\cite{oxygenate} 
Over the past three decades, the development of physically motivated descriptors, exemplified by the $d$-band center,\cite{dband1,dband2} has substantially advanced the prediction of adsorption energetics and the understanding and design of heterogeneous catalysts. 
Existing descriptors have been most successful for simple adsorbates or larger species with simple binding motifs, such as monodentate adsorption. 
However, complex adsorbates can also adopt bidentate and higher-dentate configurations, particularly on disordered working catalysts.\cite{embraceMess} 
These complexities make it challenging to disentangle the intrinsic interplay between electronic structure and adsorption geometry. 

We apply GSD to ComplexAdsorbate, a dataset of molecular adsorbates on metal surfaces that includes multidentate configurations,\cite{wwl-gpr} to test whether it can recover these coupled relationships from adsorption graphs.
For DSR, we mean-pool node and edge attributes into fixed-length tabular inputs, yielding a formula that combines electronic descriptors and edge-type statistics (Fig.~\ref{fig:case3}b,h).
For example, $\langle t_{ij}^{\mathrm{ss}}\rangle_{\mathcal{E}}\langle t_{ij}^{\mathrm{cross}}\rangle_{\mathcal{E}}$ combines surface connectivity statistics with adsorbate--surface binding abundance without resolving which molecular groups interact to which surface sites.
GSD instead combines electronic features of connected surface and adsorbate atoms before local and graph-level aggregation (Figs.~\ref{fig:case3}b,i).

The graph-derived GSD formula reveals two complementary forms of site-specific electronic--connectivity coupling (Fig.~\ref{fig:case3}i): the first descriptor ($M_1$) characterizes electronic features of connected atomic pairs, while the second ($M_2$) captures connectivity-dependent local electronic and coordination environments. 
Term~(1), $\kappa_{d,t}-r_{d,s}$, connects the neighbor's $d$-band kurtosis with the central atom's $d$-orbital radius.\cite{molecular-orbital,dband1,dband2,dband-shape} 
Term~(2), $\chi_t f_{\mathrm{d},t}$, couples electronegativity and $d$-state filling at the neighboring node, while term~(3), $f_{\mathrm{sp},t}\mathrm{EA}_s$, couples its $\mathrm{sp}$-state filling with the central atom's electron affinity. 
The retention of both $\mathrm{d}$- and $\mathrm{sp}$-state information is consistent with their roles in chemisorption models.\cite{d-sp}
The neighborhood minimum then selects the lowest combined pair score.
At a given node, the radius term shifts all candidate pair scores equally, whereas the cross-node product in term~(3) allows their ordering to change with the central atom's electron affinity. 
The selected environment therefore depends not only on the electronic properties of the connected atoms but also on their connectivity, which determines the candidate pair scores entering the neighborhood minimum.

The second descriptor combines work function $\Phi$~(4), coordination $\mathrm{CN}$~(5)--(6), and electronegativity $\chi$~(7) through central-node features and connectivity-defined neighborhood averages.
Terms~(5) and~(6) combine standardized $\mathrm{CN}_t$ with its absolute value, yielding a decreasing piecewise-linear response with a steeper slope below the training-set mean than above it.
Furthermore, the neighborhood average of $ \lvert \mathrm{CN}_s \rvert$ captures the mean absolute deviation from this reference. 
The outer maximum selects the largest absolute joint score, which may originate from either a positive or a negative local score; the response direction therefore depends on the selected node and its score's sign.
These operations allow $M_2$ to capture connectivity-dependent electronic and coordination variations obscured by global averaging.

Together, these relationships suggest that multidentate adsorption enthalpies depend on both local electronic properties and the connectivity between molecular binding groups and surface sites. 
This view echoes findings that alloys with Cu-like $d$-band centres and work functions can exhibit different multicarbon reactivity owing to surface heterogeneity and coverage distributions.\cite{peaks-pitfalls-descriptors}
Building on our earlier work on adsorption connectivity,\cite{wwl-gpr} GSD expresses the combined role of connectivity and local electronic properties in an explicit formula. The formula suggests a balance between electronic interactions across multiple binding sites and the influence of an extremal local electronic--coordination environment, offering a testable hypothesis for site-ensemble effects.
This task is particularly demanding because it requires capturing adsorbate--surface coupling across multiple binding sites and nonlinear, asymmetric responses to local coordination. 
With the current input features, distillation into two graph descriptors raises the RMSE from 0.21 to 0.66~eV, compared with 0.69~eV for DSR with $D=4$ (Fig.~\ref{fig:case3}c,k,h). 
Despite similar RMSEs, GSD shows less horizontal banding in its parity plot, consistent with retaining connectivity-dependent information in the distilled kernel.

We illustrate GSD's predictive and interpretive capabilities using tridentate CH$_2$CO adsorption on Rh(211) (Figs.~\ref{fig:case3}j--l). 
GSD predicts $\Delta E_{\mathrm{ads}}=-0.90$~eV, close to the DFT value of $-0.97$~eV, whereas DSR predicts $-0.01$~eV, giving absolute errors of $0.07$ and $0.96$~eV, respectively.
Term-level analysis shows that DSR's pooled edge-type interaction contributes $+0.58$~eV, largely cancelling negative contributions and shifting the prediction towards zero (Fig.~\ref{fig:case3}k, left). 
This term combines global edge-type averages without resolving connectivity between molecular groups and surface sites. 
GSD keeps this information through neighborhood minima, with contributions of $-4.31$ and $-1.67$~eV for Rh atoms involved in Rh--O and Rh--C binding, respectively, compared with $-0.03$~eV for surface Rh atoms associated with Rh--Rh bonds (Fig.~\ref{fig:case3}l). 
Averaging the node-level minima gives $M_1=-0.40$~eV, which combines with $M_2=+0.39$~eV and an intercept of $-0.89$~eV to yield the prediction (Fig.~\ref{fig:case3}k, right).
A systematic comparison of predictive performance for bidentate and higher-dentate adsorption configurations is presented in Table~S18. 
GSD achieves smaller absolute errors for 56 of 88 configurations with denticity $\geq 3$ (63.6\%), compared with 231 of 471 bidentate configurations (49.0\%). 
Despite similar overall RMSEs, GSD more frequently outperforms DSR for higher-dentate motifs, consistent with the benefit of retaining adsorption connectivity and local heterogeneity.

\textbf{Symbolic hierarchies across three case studies.}

Finally, the Sankey diagrams in Figs.~\ref{fig:case1}--\ref{fig:case3}g visualize the computational hierarchies retained after pruning. 
Their increasing richness across the three tasks illustrates the difficulty of compressing complex graph-structured relationships into compact formulas. 
Ribbon widths represent locally normalized flows rather than global feature importance, and large accumulated flows may reflect participation in multiple expressions. 
Importantly, the diagrams should not be read as feature-importance rankings: ribbon widths are locally normalized, and a large flow reflect participation in several intermediate expressions. 
Retaining only the largest flow at each layer could therefore omit relevant contributions and distort the learned relationships. 
Layer-wise distillation and compression preserve the hierarchies and translate them into analytical formulas. 
Thus, GSD is able to isolate readable physical routes without discarding subdominant branches required by interpretability.

\section{Discussion}\label{sec3}
Across three case studies spanning materials and chemistry, GSD reveals interpretable relationships encoded in graph-structured data that recover established physical insights while suggesting new, testable hypotheses: extremal local geometry associated with band gap variations in $\gamma$-CsPbI$_3$, local polarity–size relationships associated with hydration free energy, and site-specific electronic–-connectivity coupling in multidentate adsorption. 
Beyond predictive accuracy, the scientific value of a learned model also depends on how readily its underlying relationships can be examined and interpreted. This requires more than expressing a model in mathematical form. 
Nonlinear kernel regression can encode relationships that are difficult to interpret in terms of the original variables,\cite{kenelbad} while even compact expressions obtained by symbolic regression may contain nested nonlinear operations that hinder interpretation.\cite{srbad} 
GSD emphasizes explicit compositional structure: identifiable motifs are combined through linear aggregations within a reused hierarchy.
On the other hand, neural networks are able to capture complex dependencies but are often difficult to interpret. 
Feature-attribution methods such as SHAP\cite{shap} and attention visualizations\cite{transformer,attentionmaps} help characterize input contributions and attention patterns. In contrast, GSD distills relationships learned across the dataset into explicit formulas, uncovering shared rules for how inputs are transformed and aggregated.
In GSD, compact pruned formulas support direct human inspection, while the lengthy distilled formulas remain accessible to computer-assisted analysis.
These capabilities position GSD as an interpretable bridge between graph-based learning and scientific discovery, providing new tools for extracting explicit mathematical relationships and formulating testable hypotheses from complex graph data.

Several limitations remain. 
Most fundamentally, GSD inherits the representational ceiling of its deep teacher: it can expose relationships learned within the available graph representation and architecture search space, but cannot recover interactions that the teacher itself cannot express. 
The present search space is largely constructed from conventional local MP transformations and invariant aggregations.\cite{battaglia2018} 
Relationships that require explicit angular correlations, rotational equivariance, adaptive attention, or long-range coupling may therefore require geometry-aware GNNs or graph transformers, together with symbolic operators capable of representing their computations.\cite{transformer, sage-gnn, alignn} 
Expanding both the architecture and symbolic libraries is an important next step, although doing so will further sharpen the trade-off between predictive expressiveness and symbolic tractability.

The resulting formulas and intermediate symbolic kernels can also serve as task-adaptive graph descriptors, linking heterogeneous local environments to system-level responses while retaining In addition to supporting interpretability, explicit formulas could connect graph-based learning with large language models (LLMs), jointly assisting scientific reasoning.
Recent advances suggest that LLMs can propose mathematical hypotheses for numerical evaluation,\cite{llm-sr} while domain knowledge can guide formula discovery and interpretation.\cite{llmfeynman}
For GSD, this opens the possibility of a reciprocal process: learned graph descriptors could inform scientific hypotheses, which could in turn guide subsequent searches for structure--property relationships.
In a future GSD--LLM workflow, an LLM could draw on scientific literature to propose physically motivated operators or constraints for symbolic search.
GSD could then fit and evaluate the resulting candidate relationships on graph data.
Given feature definitions and aggregation semantics, the LLM could relate the distilled motifs to established physical models and formulate testable design hypotheses.

As autonomous scientific agents become increasingly capable of planning and executing experiments,\cite{agent} GSD could provide a descriptor-discovery module that makes the quantitative relationships informing their proposals accessible to researchers.\cite{agent-explain}
Its executable formulas could support candidate screening and help identify experiments that distinguish competing structural hypotheses.
Researchers could inspect and revise both the formulas and their physical interpretations, while new calculations or measurements test the proposed relationships and guide subsequent updates.
This integration could support a human-supervised discovery loop connecting language-based scientific reasoning with transparent, testable relationships grounded in graph data.

\section{Methods}\label{sec4}
\subsection*{Notation}

For clarity and consistency, we first summarize the graph and MP notation used throughout the GSD framework. We denote the graph topology by
\begin{equation}
\mathcal{G}=(\mathcal{V},\mathcal{E}),
\end{equation}
where $\mathcal{V}$ and $\mathcal{E}$ are the node and edge index sets, respectively. For each node $i\in\mathcal{V}$, its neighbor set is defined as
\begin{equation}
N_i=
\left\{
j\in\mathcal{V}\mid(i,j)\in\mathcal{E}
\right\}.
\end{equation}
Thus, $i$ indexes a reference atom and $j\in N_i$ indexes one of its connected neighbors.
The node-, edge-, and graph-level representations are denoted by
\begin{equation}
V=\{v_i\}_{i\in\mathcal{V}},
\qquad
E=\{e_{ij}\}_{(i,j)\in\mathcal{E}},
\qquad
u,
\end{equation}
respectively. 
The model input is written as $X=(V,E,u)$ when global features are available and as $X=(V,E)$ otherwise. 
The objective of GSD is to identify an explicit symbolic mapping
\begin{equation}
y=F(\mathcal{G},X),
\end{equation}
where $y$ denotes the target graph-level property, such as a molecular, catalytic, or materials property.

In GSD, MP flow is treated as the structural skeleton of the symbolic expression and follows the standard graph-network formalism, in which a complete block sequentially updates the edge-, node-, and graph-level representations.\cite{battaglia2018} 
The corresponding information propagation is written as

\begin{align}
e'_{ij}
&= \phi^e(e_{ij},v_i,v_j),
\label{eq:edge-update}\\
v'_i
&= \phi^v\!\left(
v_i,\rho^{e\rightarrow v}(E'_i)
\right),
\label{eq:node-update}\\
u'
&= \phi^u\!\left(
u,\rho^{e\rightarrow u}(E'),
\rho^{v\rightarrow u}(V')
\right),
\label{eq:global-update}
\end{align}
where
\begin{equation}
E'_i=\{e'_{ij}\mid j\in N_i\}
\end{equation}

collects the updated edge representations associated with node $i$, while $E'$ and $V'$ collect all updated edge and node representations, respectively. The functions $\phi^e$, $\phi^v$, and $\phi^u$ denote the edge-, node-, and graph-level update functions implemented by multilayer perceptrons (MLPs). Each $\rho$ denotes an aggregation operator selected from
\begin{equation}
\{\mathrm{sum},\mathrm{mean},\mathrm{min},\mathrm{max}\}.
\end{equation}
Building on this general formulation, GSD searches for a compact, task-specific MP flow.

The MP flow selected is instantiated and trained as a GNN predictor, hereafter referred to as the deep teacher. 
Its architecture is then fixed as the computational skeleton for SD: the order and connectivity of the learned transformations, together with the selected aggregation operations, remain unchanged. 
After training, input--output pairs are collected from each MLP component as
\begin{equation}
\label{eq:update-function}
\mathbf{y}^{(l)}
=
\phi^{t_l}\!\left(\mathbf{x}^{(l)}\right),
\qquad
t_l\in\{e,v,u\},
\end{equation}
where $l$ indexes the MLP components in the selected MP flow and $t_l$ specifies whether component $l$ performs an edge, node, or graph update. Accordingly, $\phi^{t_l}$ refers to one of the update functions $\phi^e$, $\phi^v$, or $\phi^u$ defined above. The input $\mathbf{x}^{(l)}$ contains all representations supplied to the corresponding MLP, including the outputs of the retained aggregation operators, whereas $\mathbf{y}^{(l)}$ denotes its updated representation.
The selected MP flow therefore defines an ordered set of layer-wise input--output pairs,
\begin{equation}
\left\{
\left(
\mathbf{x}^{(l)},
\mathbf{y}^{(l)}
\right)
\right\}_{l=1}^{L},
\end{equation}
which provides component-resolved supervision for SD. 
Suppressing the layer and output-channel indices, this approximation is written as
\begin{equation}
\label{eq:sr-layer-wise}
\phi(\mathbf{x})
\approx
f_{\mathrm{simple}}(\mathbf{x})
=
\sum_{k=1}^{K}
c_k\psi_k(\mathbf{x}),
\end{equation}
where $\psi_k(\mathbf{x})$ is a candidate symbolic term generated from the predefined operator set and $c_k$ is its corresponding coefficient. 
The coefficients are initially estimated using least absolute shrinkage and selection operator (LASSO) regression. Its $\ell_1$ penalty suppresses redundant symbolic terms and promotes a sparse expression.\cite{lasso}
The resulting symbolic functions are inserted back into the selected MP flow, leaving the original graph operations and aggregation pathways unchanged. 
The numerical coefficients of the assembled symbolic model are then jointly optimized against the target property. 
In this way, the distilled model reproduces both the graph-level prediction and the intermediate transformations performed by the deep teacher.
When the simple symbolic approximation is insufficient, the residual mode augments it with a linear correction:
\begin{equation}
\label{eq:sr-residual}
f_{\mathrm{res}}(\mathbf{x})
=
f_{\mathrm{simple}}(\mathbf{x})
+
\Delta f(\mathbf{x}),
\end{equation}
where
\begin{equation}
\Delta f(\mathbf{x})
=
\mathbf{W}\mathbf{x}
+
\mathbf{b}.
\end{equation}
The residual head is trained to reproduce the discrepancy between the teacher output and its symbolic approximation. It therefore keeps predictive information that cannot be readily compressed into the symbolic backbone, while the symbolic component captures the dominant analytical relationships.
In addition to the unconstrained residual model, we consider two complexity-controlled variants. 
In the $\ell_1$-regularized residual model, an $\ell_1$ penalty is applied to the residual coefficients to promote sparsity. In the targeted residual model, the residual correction is introduced only for the MP component exhibiting the largest SR fitting error. 

We next introduce the notation used to express the distilled graph-structured formulas.
Let $n$ denote the number of atoms (nodes) in a graph and $n_i$ for the number of neighbors of atom $i$.
For an oriented edge $(i,j)$, the source and destination labels used in the hierarchical kernels correspond to
\begin{equation}
x_{ij}^{\mathrm{src}}=x_i,
\qquad
x_{ij}^{\mathrm{dst}}=x_j,
\label{eq:src_dst_notation}
\end{equation}
respectively. 
Aggregation operators act either over the neighbors of an atom or over all atoms in the graph:
\begin{equation}
\begin{aligned}
\rho_e(q_i)
=
\rho_{j\in N_i}\!\left(q_{ij}\right),
\qquad
\rho_v(h)
=
\rho_{i\in\mathcal{V}}\!\left(h_i\right),
\end{aligned}
\label{eq:aggregation}
\end{equation}
Here $q_{ij}$ is an edge-level quantity and $h_i$ is a node-level quantity. 
For mean pooling, the average is taken over $\lvert N_i\rvert$ neighbors in $\rho_e$ and over $\lvert\mathcal{V}\rvert$ atoms in $\rho_v$; the other three operators need no such normalization.
These operations convert node- or edge-level quantities into local representations associated with atom $i$.

\subsection*{Constrained Message-Passing Neural Architecture Search (CMP-NAS)}
\label{subsec:cmp_nas}

The MP flow of the teacher GNN is identified using CMP-NAS. 
Rather than searching arbitrary neural-network topologies, CMP-NAS searches typed operations acting on edge-, node-, and graph-level representations according to:
\begin{equation}
E \rightarrow N \rightarrow (E \mid N)^*
\rightarrow G_{\mathrm{pool}} \rightarrow G_{\mathrm{out}},
\label{eq:mpflow_template}
\end{equation}
where $E$ and $N$ denote edge- and node-level computation slots, respectively; $(E\mid N)^*$ represents zero or more additional edge or node slots; and $G_{\mathrm{pool}}$ and $G_{\mathrm{out}}$ denote the graph-level pooling and output operations. 
Each slot corresponds to one computational unit in the MP flow, namely an MP layer. 
This constrained template ensures that every terminal architecture defines a valid graph computation and can subsequently be decomposed into component-wise input--output pairs for SD.
Meanwhile, to maintain computational tractability while ensuring that the resulting flows and symbolic expressions remain readily interpretable, we restrict the hidden dimension to at most 32 and the number of MP layers to at most 6. 
The complete operation space and model configurations, including other constraints, are provided in the SI.
The constrained architecture space is explored using MCTS with the upper confidence bound (UCT) applied to trees.\cite{uct,mcts}
A tree node $s$ represents a partially constructed MP flow, and an action $a$ appends one legal operation.
The action selected at each visited node maximizes
\begin{equation}
\operatorname{UCT}(s,a)
=
\overline{R}(s,a)
+
c_{\mathrm{uct}}
\sqrt{
\frac{\ln \sum_{a'}N(s,a')}
{N(s,a)}
},
\label{eq:uct}
\end{equation}
where $\overline{R}(s,a)$ and $N(s,a)$ are the mean reward and visit count, 
respectively, and $c_{\mathrm{uct}}=\sqrt{2}$. 
Because a partial MP flow does not define an executable GNN, the
tree-selected architectural prefix $\mathcal{A}_{1:t}=(a_1,\ldots,a_t)$ is completed after each expansion by uniformly sampling the remaining operations
$\widetilde{\mathcal{A}}_{t+1:L}=(\tilde{a}_{t+1},\ldots,\tilde{a}_L)$:
\begin{equation}
\widetilde{\mathcal{A}}_{1:L}
=
\mathcal{A}_{1:t}
\oplus
\widetilde{\mathcal{A}}_{t+1:L},
\label{eq:uniform_completion}
\end{equation}
where $L$ is the number of slots in the terminal architecture and $\oplus$ denotes sequence concatenation. 
The randomly sampled completion suffix is used only to construct and evaluate an executable terminal architecture; its operations are not inserted into the persistent search tree. 
The resulting reward is therefore backpropagated only along the selection--expansion path. 
Each completed GNN constitutes one architecture evaluation.
Terminal architecture is first assessed using a coarse validation procedure. 
The top $K_{\mathrm{fine}}$ candidates, ranked by validation RMSE and parameter count, are then retrained using five-fold cross-validation.
The architecture with the lowest merged five-fold out-of-fold RMSE is selected
as the teacher MP flow. 

The complete UCT-based procedure is summarized in Algorithm~\ref{alg:mcts_mpflow} and detailed in the SI.
We also evaluated Predictor + Upper Confidence Bound applied to Trees (PUCT), in which the policy prior was trained using architectures sampled under the current scheme.\cite{puct,alphago,ND2} 
Because PUCT provided no consistent performance gain while introducing an additional training stage and greater workflow complexity, we thus use UCT in the final implementation (see the SI).

\begin{algorithm}[t]
\caption{UCT-MCTS for constrained message-passing architecture search}
\label{alg:mcts_mpflow}
\begin{algorithmic}[1]
\Require $N_{\mathrm{eval}}$, $c_{\mathrm{uct}}$, and $K_{\mathrm{fine}}$
\State Initialize the root with an empty MP flow
\For{$k=1,\ldots,N_{\mathrm{eval}}$}
    \State Select a partial flow using UCT and expand one legal action
    \State Uniformly complete the remaining slots
    \State Train and evaluate the resulting terminal GNN
    \State Backpropagate its reward along the tree path
\EndFor
\State Re-evaluate the top $K_{\mathrm{fine}}$ flows by five-fold cross-validation
\State \Return the flow with the lowest merged out-of-fold RMSE
\end{algorithmic}
\end{algorithm}

\subsection*{Symbolic Regression Setup}
\label{method-3}
We employ TorchSISSO,\cite{torchsisso} a native PyTorch implementation of the Sure Independence Screening and Sparsifying Operator (SISSO) framework,\cite{sisso} as the SR engine for both DSR and GSD.
For a controlled comparison, both approaches use the same operator set, comprising binary operations $\{+,-,\times,/\}$ and unary transforms $\{x^{2},x^{-1},e^{-x},\lvert x \rvert,\sqrt{x}\}$.
The feature-expansion depth is $n_{\mathrm{exp}}=2$, termed the \emph{rung} in SISSO and controlled by \texttt{n\_expansion} in TorchSISSO.
Each SR keeps at most $n_{\mathrm{term}}=D=4$ descriptor components (\texttt{n\_term} in TorchSISSO; \texttt{desc\_dim} in SISSO).
The screening size is $k=15$ features per SIS iteration (\texttt{k} in TorchSISSO; \texttt{subs\_sis}, renamed \texttt{nf\_sis} in newer SISSO versions).
Datasets are partitioned at the sample level into training, validation, and test sets with an \(80\%/10\%/10\%\) split. 
In addition, for layer-wise SD, an optional Lasso-based preselection retains at most \(10\) input features and caps the number of training rows at \(5000\) before the SISSO search.

\subsection*{Formula Finder Workflow}
\label{method-4}

Those selected Targ. and $\ell_{1}$ Res. models serve as inputs to the formula finder workflow, and are compressed into compact, feature-only formulas (see Fig.~\ref{fig:mcts}d)

The composed symbolic network is first printed as a hierarchical kernel with two blocks: a graph-level readout that pools a small set of placeholders, and a set of nested definitions for those placeholders.
Node/graph bodies are denoted as NB and edge-aggregation bodies EB.
Four controls govern this print---thresholds on symbolic and residual coefficients, and caps on the numbers of pooling branches and terms per definition, and candidate kernels are ranked after the outer coefficients are refit on the training split.

The selected kernel is then refined with its operators, exponents, and hierarchy held fixed, while its internal numerical coefficients and linear readout are optimized.
The resulting kernel teacher guides later compression; note that it is not the deep GNN teacher used in SD.

Feature-only leaves of this kernel, together with the aggregation path from each leaf to the graph-level output, define candidate descriptors that we call motifs (M).
A greedy scan on validation accuracy, optionally regularized by agreement with the kernel teacher, retains a small motif set of size $k$.
The outer coefficients are then refit to give
\begin{equation}
    \hat{y}(X)=b+\sum_{i=1}^{k}c_i M_i(X),
    \label{eq:formula_finder_outer}
\end{equation}
where each $M_i$ is an explicit combination of input features and graph operations.
Detailed procedures and hyperparameters are given in the SI, Section~S5.

\subsection*{Comparison of GSD and DSR}
\label{method-5}

For comparison, the DSR (SISSO) baseline directly fits the target with the same settings as SR within SD: $\mathrm{rung}=2$, $D=4$, and a screening size of 15 features.
Figs.~\ref{fig:case1}--\ref{fig:case3} exhibit the predictive performance of DSR and GSD across three case studies.

DSR uses task-specific fixed-length inputs: the eight predefined tabular structural features from the original $\gamma$-CsPbI$_3$ study,\cite{metastable} the available global molecular descriptors for FreeSolv, and mean-pooled node and edge features for ComplexAdsorbate, which lacks native graph-level features. 
Across all three tasks, GSD operates directly on node and edge features, together with native graph-level features when available; the eight predefined structural descriptors for $\gamma$-CsPbI$_3$ are excluded from its inputs. 
Complete feature definitions and graph-construction procedures are provided in Tables~S1--S3.

The simpler $\gamma$-CsPbI$_3$ and FreeSolv tasks require only modest $k$ and $D$; their comparisons are provided in Section~S7 of the SI. 
Fig.~S2 compare the predictive accuracy and computational cost of GSD and DSR as model size varies for case 1 and 2. 
In these two easier tasks, the available expert-defined descriptors already encode informative relationships and therefore provide a well-established basis for direct SISSO. 
GSD does not yield a substantial improvement in predictive accuracy for these two cases, nevertheless recovers these relationships directly from elementary node and edge features, without relying on the predefined descriptors. 

We therefore focus the detailed comparison between GSD and DSR on ComplexAdsorbate, where predictive accuracy and computational cost are evaluated as a function of $k$ and $D$ in Fig.~\ref{fig:mcts}e,f. 
The results demonstrate a more favorable fidelity--computational cost trade-off for GSD in this more challenging ComplexAdsorbate task (see Methods for details).
SISSO is particularly effective when informative input features admit a compact representation: nonlinear feature construction followed by screening and sparse selection provides substantial fitting flexibility with only a few descriptor components.\cite{sisso} 
GSD itself exploits this capability in its component-wise SR. 
However, extending direct SISSO to higher descriptor dimensions becomes computationally demanding. 
At rung 2 and $D=7$, the calculation requires approximately 12~h on 128 CPU cores.
For GSD, constructing the distilled teacher kernel takes approximately 3.6~h on a single NVIDIA RTX 4060 GPU. 
This kernel is subsequently reused to construct formulas with different numbers of kept motifs, $k$, with little additional computation. 
The final formula complexity is therefore largely decoupled from the dominant cost of distillation. 
Retaining more motifs improves accuracy while leaving runtime nearly unchanged: GSD reaches an overall RMSE of approximately 0.47~eV at $k=30$, compared with approximately 0.57~eV for SISSO at $D=7$. 
Under the reported hardware and formula-construction settings, GSD thus achieves higher accuracy within a shorter runtime.

On the other hand, algebraic compactness does not necessarily imply physical interpretability. 
A short expression can remain difficult to interpret when nested nonlinear operators obscure the relationships among its inputs. 
GSD organizes its prediction as an additive combination of graph motifs (Eq.~\ref{eq:formula_finder_outer}), each with explicit feature transformations and aggregation operations. 
Although this graph notation can lengthen the displayed formula, it makes the construction and contribution of each descriptor traceable. 
At fixed motif complexity, the expression grows linearly with $k$, allowing model detail to be increased while retaining an explicit account of how structural information enters the prediction.

\backmatter

\bmhead{Data availability}
Graph representations of all three cases used in this work, provided in PyTorch Geometric (PyG) format, will be made public on GitHub upon acceptance.

\bmhead{Code availability}
The source codes for SISSO and TorchSISSO are publicly available at \url{https://github.com/rouyang2017/SISSO} and \url{https://github.com/PaulsonLab/TorchSISSO}, respectively. 
The source code of GSD will be made public on GitHub upon acceptance.

\bmhead{Supplementary information}
Additional details on database, primary features, machine learning methods are provided.

\bmhead{Acknowledgments}
The authors gratefully acknowledge support from the Max Planck Computing and Data Facility (MPCDF). W.X. acknowledges funding from National Natural Science Foundation of China (NSFC). The authors thank Hongzhi Shi for fruitful discussions.

\bmhead{Author contributions}
W.X. and Y.Y. conceived and conceptualized the project. Y.Y. carried out the methods development and data analysis. W.X. supervised the project. Y.Y. wrote the manuscript. W.X., and K.R. revised the manuscript. 

\bmhead{Competing interests}
The authors declare no competing interests.

\bibliography{bibliography}
\end{document}